\documentclass[aps, prb, twocolumn, superscriptaddress , showpacs]{revtex4-1}

\usepackage{graphicx}
\usepackage{dcolumn}

\input colordvi

\begin{document}

\title{Importance of High Angular-Momentum Channels in Pseudopotentials for Quantum Monte Carlo}
\author{William W. Tipton}
\affiliation{Department of Materials Science and Engineering,
                  Cornell University, Ithaca, New York 14853, U.S.A.}
\author{Neil D. Drummond}
\affiliation{Department of Physics,
                  Lancaster University, Lancaster LA1 4YB, United Kingdom}
\affiliation{TCM Group, Cavendish Laboratory, 
                  University of Cambridge, J. J. Thomson Avenue, Cambridge CB3 0HE, United Kingdom}
\author{Richard G. Hennig}
\affiliation{Department of Materials Science and Engineering,
                  Cornell University, Ithaca, New York 14853, U.S.A.}
\date{\today }

\begin{abstract} Quantum Monte Carlo methods provide in principle an accurate treatment of the many-body problem of the ground and excited states of condensed systems.  In practice, however, uncontrolled errors such as those arising from the fixed-node and pseudopotential approximations often limit the quality of results.  We show that the accuracy of quantum Monte Carlo calculations is limited by using available pseudopotentials.  In particular, it is necessary to include angular momentum channels in the pseudopotential for excited angular momentum states and to choose the local channel appropriately to obtain accurate results.  Variational and diffusion Monte Carlo calculations for Zn, O, and Si atoms and ions demonstrate that these issues can affect total energies by up to several eV for common pseudopotentials.  Adding higher-angular momentum channels into the pseudopotential description reduces such errors drastically without a significant increase in computational cost. \end{abstract}


\maketitle

\section{Introduction} 

Computational electronic structure methods have been extremely useful in developing our understanding of the atomic and electronic structures of real materials.  As methods have become more accurate and their implementations increasingly efficient, simulation and calculation have taken some of the burden of finding and characterizing new materials off experimental work.\cite{kohn_nobel}

Density functional theory (DFT), in particular, has been widely applied to many systems in recent decades. It is computationally efficient compared to other methods with similar accuracy, and robust, user-friendly software packages have made the method easy to apply.  However, its accuracy is still insufficient for some applications, and the lack of a systematic way to improve its results or estimate its errors has hampered progress.  In particular, electron correlation effects can be significant in many complex materials and are not captured accurately by many of the commonly-used density functionals.  The development of functionals which accurately describe the electronic band gap, van der Waals interactions, and other electronic properties of materials is still an active area of research.\cite{anisimov_ldau,PhysRevLett.80.4153,heyd:8207, Dion92}

These issues are overcome by methods which treat quantum many-body effects explicitly from the outset such as quantum Monte Carlo (QMC).  QMC methods are among the most accurate many-body methods and  can reliably and accurately predict ground-state expectation values for many systems and have often been used as a benchmark for DFT work.\cite{foulkes}  Among the quantum Monte Carlo methods, variational Monte Carlo (VMC), diffusion Monte Carlo (DMC), and auxillary-field quantum Monte Carlo\cite{PhysRevB.75.245123} are the most mature in terms of applicability to solid state systems.  We treat only VMC and DMC in this work and refer to them collectively as QMC.

As computers become faster and high-quality software packages for QMC such as CASINO\cite{casino}, QMCPACK\cite{qmcpack}, QWALK\cite{Wagner20093390}, and CHAMP\cite{champ} mature, these calculations are becoming less challenging.  It is therefore important to identify and propagate the best-practice procedures for performing these calculations as they become more routine.

QMC and other correlated-electron methods, particularly for heavy elements, usually employ the pseudopotential approximation to reduce the computational cost.  The common form is the non-local, norm-conserving pseudopotential\cite{PhysRevLett.43.1494} which includes different potentials for each angular-momentum component of the wavefunction.

In this work, we determine the error in the energy due to an insufficient number of angular-momentum channels in the pseudopotential and discuss other sources of error in QMC calculations.  We show that pseudopotentials which include channels to account for higher angular-momentum components of the wavefunction are necessary for performing accurate pseudopotential calculations in QMC.  Such pseudopotentials are not the norm in the literature, and we suggest that this be corrected in order that QMC methods be suitable for routine applications to scientifically and technologically interesting systems.

\section{Background} 

The computational cost of all-electron QMC scales approximately as $Z^{5.5}$ or $Z^{6.5}$ with respect to the atomic number.\cite{ceperley86,hammond87}  This scaling makes the direct application of all-electron QMC to heavy atoms difficult.  In practice, many properties of atoms are primarily due to the behavior of and interactions between valence electrons, and so a pseudopotential approximation is commonly used to remove core electrons from the calculation and reduce the necessary computational effort.

Modern pseudopotentials are non-local in the sense that they act differently on distinct angular-momentum components of the wavefunction. This is necessary to accurately capture the effects of the nucleus and core electrons on the valence electrons since the pseudopotential not only represents the effective electrostatic potential but also enforces orthogonality of the valence orbitals to the lower-energy states of the same angular momentum.  

Of course, there is no clear distinction between core and valence electrons in many-body methods as the electrons are indistinguishable particles.  Thus, the application of pseudopotentials does neglect exchange and correlation interactions between the valence and removed core electrons as well as that between the core electrons themselves.  These errors are not explicitly accounted for in the calculations.  However, the core-core interactions largely cancel out when considering energy differences, and the core-valence interactions may be kept small by a choice of core size which leads to significant spatial separation between the core and valence electron densities.  These techniques can lead to results as accurate as all-electron calculations.\cite{hammond87}  Additionally, the use of core-polarization potentials can account for some core-valence correlations.\cite{muller:3297,PhysRevB.47.15413,PhysRevB.67.035121}

To introduce non-local pseudopotentials in QMC, the electron-ion potential of any given atom is divided into a local potential $\hat{V}_{\textrm{loc}}$ which is applied to the whole wavefunction and several corrections $\hat{V}_{\textrm{nl}}$ which account for the difference between the local potential and those which should be seen by the individual angular-momentum components of a wavefunction:
\begin{equation}
\hat{V}_{\textrm{loc}}+ \hat{V}_{\textrm{nl}} = \sum_{i=1}^{N_\mathrm{el}} V^{\textrm{ps}}_{\textrm{loc}}(r_i) + \sum_{i=1}^{N_\mathrm{el}} \hat{V}^{\textrm{ps}}_{\textrm{nl},i}.
\label{eqn:pps}
\end{equation}
The nonlocal operator $\hat{V}^{\textrm{ps}}_{\textrm{nl},i}$ acts on a function $f({\bf r}_i)$ by
\begin{equation}
\label{eqn:non_local_op}
\hat{V}^{\textrm{ps}}_{\textrm{nl},i}f({\bf r}_i) = \sum_{l,m} V^{\textrm{ps}}_{\textrm{nl},l} (r_i) Y_{lm}(\Omega_i) \int_{4\pi}Y^*_{lm}(\Omega'_i) f({\bf r}'_i) d\Omega'_i \,,
\end{equation}
where the angular integral in the operator projects the wavefunction onto spherical harmonics.  Each angular-momentum component thus ``feels'' its own potential, $V^{\textrm{ps}}_{\textrm{nl},l}(r)$, which is only a function of the electron-nuclear distance $r$ and accounts for the difference between the desired $l$-dependent potential and the local channel $V^{\textrm{ps}}_{\textrm{loc}}$.  The local potential, or local channel,  $V^{\textrm{ps}}_{\textrm{loc}}(r)$, is by convention chosen to be the exact potential applied to one of the angular-momentum components, and so sum in Eq.~(\ref{eqn:non_local_op}) need not include this local component.\cite{foulkes}

The choice of local channel itself is arbitrary but is often chosen for convenience during the pseudopotential design process.  In particular, judicous choice of the local channel is often necessary to avoid the problem of ghost states which can arise due to the Kleinman-Bylander transformation.\cite{PhysRevB.41.12264}  We will see that the same choice of local channel that is suitable for that transformation may not be the best with regards to accuracy of QMC calculations.\cite{PhysRevB.78.245124}

In an independent electron theory such as Hartree-Fock or DFT, atomic wavefunctions are composed of some number of the lowest-energy single-particle orbitals.  For example, in these frameworks, the electronic configuration of an oxygen atom may be written as $1s^22s^22p^4$.  Notice that this wavefunction contains no angular-momentum components above $l=1$.  Thus, a non-local pseudopotential in the above form which acts on these single-atom wavefunctions need not include terms $V^{\textrm{ps}}_{\textrm{nl},l}$  for $l \ge 2$ if it is to be used to calculate ground-state atomic properties.

The situation is not so simple in the case of solids and other extended systems where changes in the wavefunctions due to bonding effectively introduce higher angular-momentum components.  Indeed, in the case of systems such as Si and other second row elements, one may find wavefunctions with significant higher angular-momentum character. In this case, it may be necessary to use a pseudopotential with a $d$-channel when studying these systems in DFT, especially in the high-pressure regime.  However, these errors often cancel when considering energy differences and are frequently neglected in practice.\cite{PhysRevB.40.2980,PhysRevB.64.045119}

In QMC and other correlated-electron methods, excitations of the wavefunction into higher angular-momentum states arise immediately, {\it i.e.}, even for atoms.  In VMC, wavefunctions may be represented by the product of a Slater determinant of single particle orbitals and the so-called Jastrow factor.  The Jastrow is a positive function of inter-particle distances, and its purpose is to directly account for many-body correlation effects.  Naturally, the VMC wavefunction is then no longer entirely composed of the lowest-order spherical harmonics.  The situation in DMC (as well as other correlated-electron methods) is analogous.\cite{foulkes}

Notice from Eq.~(\ref{eqn:pps}) that, in the absence of a pseudopotential channel to deal with the higher-angular momentum components of the wavefunction, these components simply feel the local channel.  This is incorrect and may be drastically so, especially in the case where the local channel was designed to enforce orthogonality to the lower-energy orbitals with a particular angular momentum.  This can lead to sizeable errors in total energy calculations.

Now, this effect is not a particularly surprising one and certainly has been understood by some in the density functional theory community since the beginnings of the use of pseudopotentials in that field (see, for example, Ref.~\onlinecite{PhysRevB.18.5449}). However, inclusion of so-called higher angular-momentum channels is not the normal practice in the development of potentials for use with QMC.

There are a limited number of pseudopotentials available for use with QMC.  In particular, the application of projector-augmented waves\cite{PhysRevB.50.17953} or ultra-soft pseudopotential\cite{PhysRevB.41.7892} techniques in QMC is currently not feasible since the DMC operators for the projectors and the augmentation charge are unknown.  However a number of semi-local pseudopotentials have been developed with QMC applications in mind.  Greeff {\it et al.} developed a carbon pseudopotential which included $s$- and $p$-channels.\cite{greeff:1607}  Ovcharenko {\it et al.} applied a similar methodology to produce pseudopotentials for Be to Ne and Al to Ar with $l_{\textrm{max}}=1$.\cite{ovcharenko:7790}  Burkatzki {\it et al.} present potentials for many of the main group elements \cite{burkatzki:234105} and for the 3$d$ transition metals.\cite{burkatzki08}  Their Si and Zn potentials have 3 channels, and their O potential has 2.  These authors all cite the rule of thumb that $l_{\textrm{max}}$ should be at least as high as the highest angular-momentum component in atomic core.  Trail {\it et al.} developed a variety of pseudopotentials for all elements from H to Hg.  These all have exactly 3 channels and are associated with the CASINO code which, until recently, only supported pseudopotentials with exactly 3 channels.\cite{trail05}

\section{Methodology} 

We determine how the number of channels and the choice of local channel affects the energy for several atoms and ions.  We compute the total energies and first and second ionization energies of the zinc, oxygen, and silicon atoms using several related pseudopotentials. These elements provide interesting test cases due to their varied electronic structures. Additionally, we are interested in the application of QMC methods to bulk semiconductors such as Si and ZnO.~\cite{Hennig10, Parker11}

The oxygen and silicon pseudopotentials are based on those by Driver {\it et al.}\cite{driver10}, and the zinc pseudopotential is based on that by Bennett and Rappe.\cite{rappe_pps}  All three are generated using the Opium pseudopotential code.\cite{opium}  Cutoff radii and basis functions for the construction of the pseudopotentials were chosen to minimize the difference in pseudopotential and all-electron valence energy levels calculated in DFT using the PBE exchange-correlation functional~\cite{PBE}  for several electronic configurations.

\begin{table}[t]
\caption{Choices of angular-momentum channels and local channels for the various pseudopotentials considered for oxygen, silicon and zinc.}
\label{tbl:potls}
\begin{ruledtabular}
\begin{tabular}{l c  c  c  c}
  & \multicolumn{2}{c}{\bf Standard} & \multicolumn{2}{c}{\bf Augmented} \\
 & {\bf Channels} & {\bf Local} & {\bf Channels} & {\bf Local} \\
\colrule
{\bf O} & $s$, $p$ & $s$ or $p$  & $s$, $p$, $d$  & $s$ or $p$  \\
{\bf Si} & $s$, $p$ & $s$ or $p$  & $s$, $p$, $d$  & $s$ or $p$  \\
{\bf Zn} & $s$, $p$, $d$ & $s$ or $d$  & $s$, $p$, $d$, $f$  & $s$, $d$, or $f$  
\end{tabular}
\end{ruledtabular}
\end{table}

\begin{figure}
  \centering
\includegraphics[width=8cm]{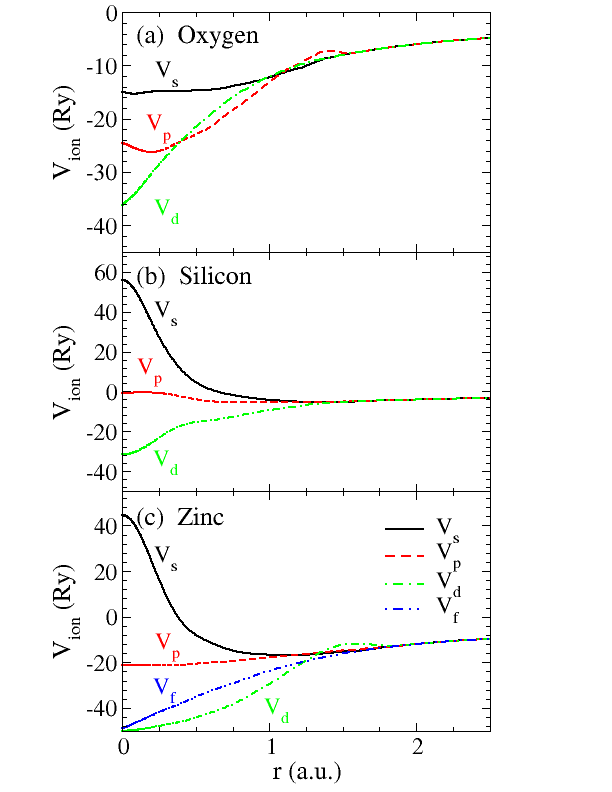}
\caption{(color online) Pseudopotentials for O, Si, and Zn.}
\label{fig:pps}
\end{figure}

Table~\ref{tbl:potls} lists the angular-momentum channels and the choice of local channel for each of our pseudopotentials.  For each element, we consider (i) pseudopotentials with the minimum number of channels ($s$ and $p$ for Si and O; $s$, $p$ and $d$ for Zn) and (ii) pseudopotentials that contain an additional angular-momentum channel ($d$ for O and Si; $f$ for Zn).  We refer to the first set as \emph{standard} pseudopotentials and the second one as \emph{augmented} pseudopotentials.  For the local channel we consider the $s$ and $p$ channels for Si and O and the $s$, $d$ or $f$ channels for Zn. This results in a total of 13 pseudopotentials as listed in Table~\ref{tbl:potls}.  Figure~\ref{fig:pps} shows the distance dependence of the angular momentum channels for the various pseudopotentials.  For Si and Zn, we confirmed that the pseudopotentials accurately describe the lattice parameters of the ground state crystal structure and for O, we confirmed that the pseudopotential reproduces the dimer bond length at the DFT level.

QMC calculations were performed using the CASINO code.\cite{casino}  We implemented support for pseudopotentials with an arbitrary number of angular-momentum channels in CASINO.  Total energy calculations are performed on the 9 isolated ions with each of the applicable pseudopotentials.  The VMC calculations used Slater-Jastrow variational wavefunctions with orbitals expressed in a blip basis.\cite{PhysRevB.70.161101}  The single-particle orbitals were generated using the PWSCF code~\cite{pwscf} and the PBE exchange-correlation functional.~\cite{PBE} Plane-wave cutoffs of 70~Ry for oxygen and silicon and 100~Ry for zinc were used to converge the total energies to 2~meV.  The known magnetic states of the atoms and ions were used.  The Jastrow factor is a non-negative function of inter-particle distances and includes two-body electron-electron and electron-nucleus and three-body electron-electron-nucleus terms as implemented in CASINO.\cite{PhysRevB.70.235119} 
The backflow transformation\cite{PhysRevE.74.066701} was not found to provide any significant benefit in these cases.  Parameters were added to the Jastrow factor of the trial wavefunction gradually during its optimization.  The Jastrow parameters were optimized using variance minimization~\cite{Umrigar88} followed by energy minimization in the final step.~\cite{Umrigar06} Trial wavefunctions were evaluated by their mean energy plus three times the statistical error in the energy, following Ref.~\onlinecite{toulouse:084102}.

Several additional details of our VMC calculations are noteworthy.  First, the integral in Equation~(\ref{eqn:non_local_op}) is performed on a spherical grid in real space.  This integration mesh must be chosen to be sufficiently dense to accurately calculate the contributions to the energy from higher angular-momentum compontents of the wavefunction and thus evaluating the effects which are the focus of this paper.  Secondly, it is the default behavior of the CASINO code that the non-local contributions to the energy are assumed constant and are not re-calculated during a variance minimization step.  In many systems, this improves the runtime of the algorithm significantly while still giving good results --- in some cases it actually improves the performance of the variance minimization.  However, as we will see, the non-local contributions are significant in many of our calculations. We found it necessary in many cases to re-calculate the non-local contributions to the energy at each step of the optimization to ensure the stability of the optimization process during Jastrow optimization.

Our DMC calculations were performed using the pseudopotential locality approximation.~\cite{mitas:3467} For each system, we performed $1,000$ equilibration and $3,000$ accumulation steps on each of $256$ processors with a timestep of $0.01$ Ha$^{-1}$ and a target population of $2,000$ walkers.  

Finally, atomic ionization energies are simply differences between the total energies of the appropriate species.


\section{Results and Discussion} 

\begin{figure}
\includegraphics[width=8.5cm]{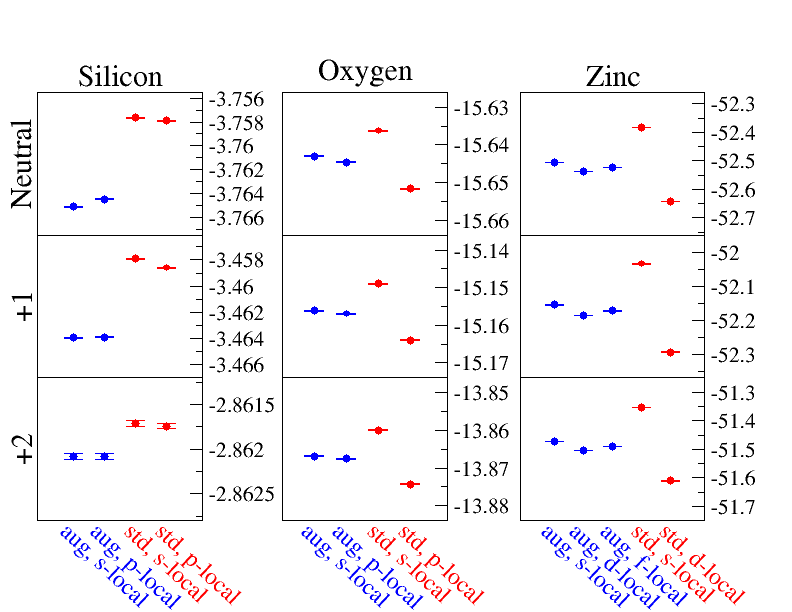}
\caption{(color online) Total VMC energy in Hartree and statistical error in the energy of each species with respect to each Hamiltonian.  Pseudopotentials are denoted according to the choice of local channel and as `aug' if they are augmented with an additional channel or `std' otherwise.}
\label{fig:vmc_plot}
\end{figure}

\begin{figure}
\includegraphics[width=8.5cm]{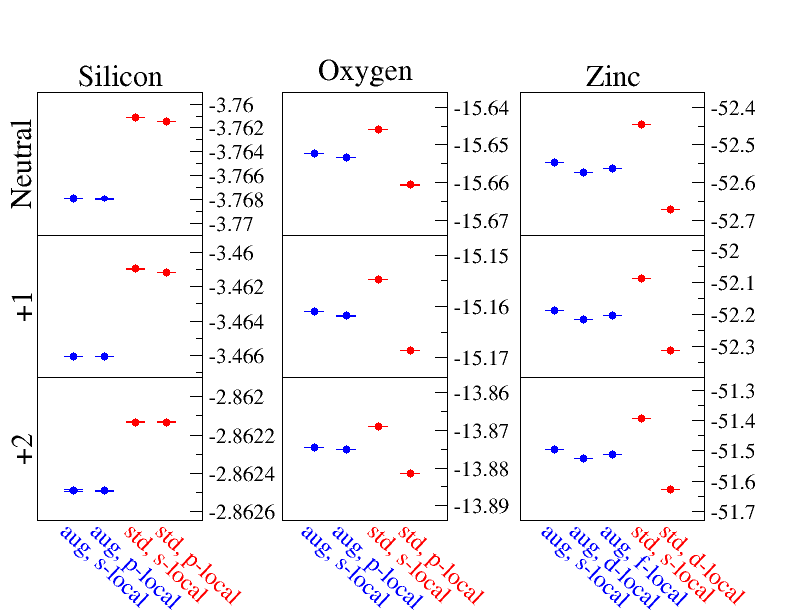}
\caption{(color online) Total DMC energy in Hartree and statistical error in the energy of each species with respect to each Hamiltonian.  Pseudopotentials are labeled as in Fig.~\ref{fig:vmc_plot}.}
\label{fig:dmc_plot}
\end{figure}

Figures \ref{fig:vmc_plot} and \ref{fig:dmc_plot} show the energies for each ion-pseudopotential combination for VMC and DMC, respectively.  The error bars indicate only the statistical uncertainties in the energies associated with the QMC calculations.  First, it is important to notice the axes.  The variation in the total energies differs between the species.  For Si, it is on the order of milli-Hartrees, while for Zn, it is on the order of tenths of Hartrees.  O falls somewhere in between.

\begin{figure}
\includegraphics[width=8.5cm]{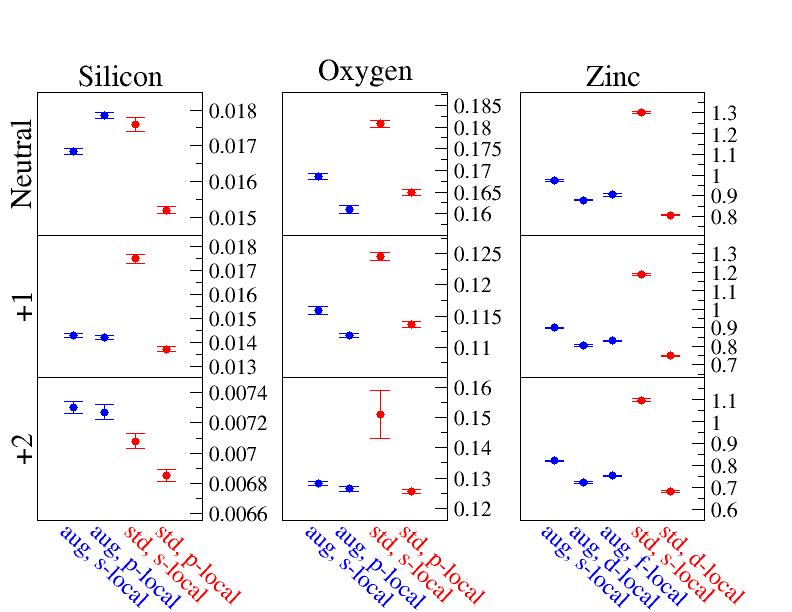}
\caption{Variance in local energy in atomic units and associated statistical error of each VMC calculation.  Hamiltonians are labeled as in Figure \ref{fig:vmc_plot}.}
\label{fig:vmc_vars_plot}
\end{figure}

\begin{figure}
\includegraphics[width=0.5\textwidth]{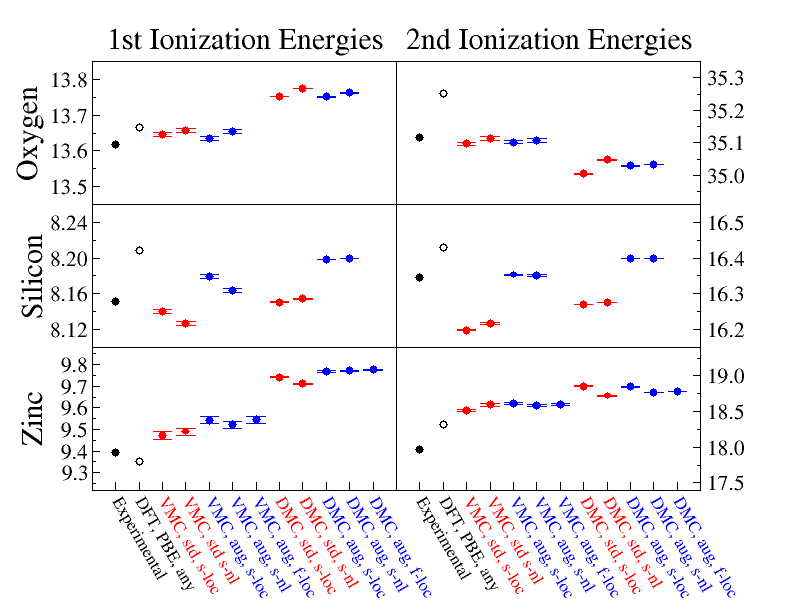}
\caption{Comparison of the ionization energies in eV for oxygen, silicon and zinc in DFT, VMC and DMC for the different choices of pseudopotential with experiment.  Energies for oxygen and silicon include corrections for the pseudopotential error at the DFT level as described in the text.}
\label{fig:ion_en_plot}
\end{figure}


Consider now the VMC and DMC total energies.  The two sets of data exhibit similar trends which is to be expected since they arise from the differences in the pseudopotentials.  The first thing to notice is that the calculations using an augmented pseudopotential ({\it i.e.}, the two or three left-most data points in each panel) are largely in agreement with each other, while this is not the case for the standard potentials.  That is, the choice of local channel has a large effect on the total energy when using the standard pseudopotentials since the higher-$l$ components of the wavefunction see that local channel.  When using the augmented potentials, more of the wavefunction sees its correct potential, and the choice of local channel has less effect on the result of the calculation.

Indeed, if we take the calculations with augmented potentials to indicate the correct result, we can understand the errors in the other total energies in terms of which potentials are incorrectly applied to certain components of the wavefunction.  As seen in Figure \ref{fig:pps}, the $s$-channel is the most repulsive for each of the species, the $p$-channel is in the middle, and the $d$-channel is the most attractive. The $f$-channel is slightly above $d$-channel in the case of zinc.

Thus, we expect that calculations in which components of the wavefunction incorrectly see the $s$-channel to be too high in energy.  Indeed these data points (which are the second right-most point in each frame of the total energy plots) exhibit this trend.  Similarly, the right-most data point in each frame corresponds to a calculation wherein any higher $l$ component of the wavefunction sees the $d$-channel, and these results are erroneously low in energy.  Even the residual differences between the energies calculated using the augmented potentials follow this trend. This is indicative of small amounts of yet-higher $l$ character in the wavefunctions.

We can get some idea of the amount of virtual excitations that might be present in the many-body ground state of each of the species by considering the energies of these excitations.  In this way, we can understand the expected magnitude of these effects. 


\begin{table}
\renewcommand{\tabcolsep}{3pt}
\caption{Lowest-energy excitations in eV to higher-$l$ states for each species from experiment.\cite{sugar:1803,martin:323,o_crc_spectra_93,nist_spectra}  }
\label{tbl:excitations}
\begin{ruledtabular}
\begin{tabular}{l d d d  }
 {\bf Species}  & {\bf O} & {\bf Si} & {\bf Zn} \\
\colrule
{\bf Neutral} & 12.08 & 5.86 & 8.53 \\
{\bf Singly-Ionized} & 28.7 &  9.84 & 14.54 \\
{\bf Doubly-Ionized} & 40.23 &  17.72 & 31.9 \\
\end{tabular}
\end{ruledtabular}
\end{table}

Table~\ref{tbl:excitations} presents the lowest energy excitation to a higher angular-momentum state for each of the three atoms for the various charge states.  The excitation energies of these states increase with the level of ionization.  Additionally, the $d$ levels are relatively high in oxygen but low in silicon.  The $f$ levels in Zn trend in between.  Thus, we expect that the effects in total energy described in this paper will be especially pronounced for the neutral species relative to the positive ones and for silicon relative to oxygen.  Note that this effect due to the lower excitation energies here has implications not only for the atomic wavefunctions.  Lower-energy states are more likely to participate in bonding in molecules and solids, and it is especially important to design pseudopotentials to account for that.

Indeed, for the silicon species, the decreasing significance of the extra channel with increasing charge is clear.  This effect is less readily apparent in oxygen and zinc data, and is likely obscured by another notable and correlated effect.  The variance in the local energy for each VMC calculation is shown in Figure \ref{fig:vmc_vars_plot}.  Eigenstates of the Hamiltonian have a variance of zero, and higher variances indicate worse approximations of the ground-state wavefunction.  That is, higher variances are correlated with higher total energies.  From the figure, it is clear that our calculations using the standard pseudopotentials with the $s$-channel local resulted in poorer-quality wavefunctions.  These higher variances are not fundamental properties of the system but demonstrate that it is more difficult to optimize a Slater-Jastrow wavefunction with respect to these less realistic Hamiltonians.  Additionally, the wavefunctions might be better described by a multi-determinant expansion, especially in the case of oxygen.


Finally, the first and second ionization energies for each element are shown in Figure \ref{fig:ion_en_plot}.  The DFT ionization energies were computed from the same calculations used to create the orbitals for the VMC trial wavefunction, and finite-size effects due to periodic boundary conditions were treated using the method of Makov and Payne.\cite{PhysRevB.51.4014} 
Several sources of errors may explain the deviation of these results from experimental values.  
First, we calculated spin-orbit corrections to the total energies at the DFT level and found that they largely cancel in the ionization energies, resulting in corrections too small to account for the observed differences in the ionization energies between QMC, DFT and experiment.

Second, the pseudopotential approximation itself leads to several errors other than those focused on in this paper.  By removing explicit treatment of core electrons from the calculation, we are neglecting correlations between the core and valence electrons.  This is minimized but not altogether eliminated by designing pseudopotentials so that the core and valence electrons are spatially separated. The core-valence correlation may be particularly important for the case of zinc where the $3d$ valence electron state have a sizeable spatial overlap with the $3p$ core electron states and may explain the large errors in the ionization energies.  Thirdly, evaluation of the pseudopotentials in DMC is subject to the locality approximation~\cite{mitas:3467} used in this work or the lattice-regularized method by  Casula.\cite{PhysRevB.74.161102}  Pozzo and Alf\`e~\cite{Pozzo08} found that, in magnesium and magnesium hydride, the errors of the locality approximation and the lattice-regularized method are comparably small, but that the lattice method requires a much smaller DMC time step.

Finally, our pseudopotentials themselves could likely be further optimized within the same framework.  In particular, the scattering properties of PBE pseudopotentials may be poor at certain energy scales, and HF potentials might perform better in conjunction with correlated-electron methods.\cite{greeff:1607}  To test the accuracy of the pseudopotential approximation for oxygen, we performed an all-electron, single-determinant DMC calculation of an isolated oxygen atom with a Slater-Jastrow trial wavefunction and found an ionization energy of 13.611(7) eV, in close agreement with the experimental value.  This strongly suggests that the errors in the DMC ionization energies in Fig.~\ref{fig:ion_en_plot} are due to the pseudopotentials.

With this in mind, we estimated corrections for the pseudopotential error at the DFT level for oxygen and silicon.  All-electron DFT/PBE ionization energies were calculated using the Gaussian code\cite{g09} converged with respect to the atomic basis.  The difference between these ionization energies and those found with PBE using pseudopotentials should capture much of the error due to the pseudopotentials, and we have added these differences to the QMC results for oxygen and silicon shown in Figure \ref{fig:ion_en_plot}.

In the case of zinc, the error in the ionization energy in QMC stems from the poor description of the $3d$-levels of the zinc atom in DFT.  Semilocal functionals are known to place the $3d$ level of the Zn atom significantly too high \cite{PhysRevB.74.245115, Gunnarson79}.
This results in an incorrect description of the $d$-channel of the pseudopotential and of the $3d$-orbital in the trial wave function which is reflected in both the large energy variance and large deviation of the QMC ionization energy from experiment.  Correcting for this error at the DFT level by adding the energy difference between an all-electron and pseudopotential DFT calculation to the QMC results is found to be insufficent.


\section{Conclusions} 
We showed that pseudopotentials which include channels to account for higher angular-momentum components of the wavefunction are necessary for performing accurate pseudopotential calculations in QMC.  For O, Si and Zn, we determined how the number of angular-momentum channels and the choice of local channel in the pseudopotential affects the total energy and ionization energies of these atoms in QMC.  We find a sizable error in the total energies for any choice of local channel when the pseudopotentials do not include at least one additional angular-momentum channel above the highest angular-momentum component of the ground state wavefunction of the atom.
This is because, contrary to single-electron mean-field methods such as DFT and HF, atomic ground state wavefunctions in correlated-electron methods include higher angular-momentum character.  These components effectively see the wrong potentials when using standard pseudopotentials.  This situation is expected to be even more pronounced in the case of solids and molecules.

Our results suggest that the best practice is to include at least one channel in the pseudopotential above the highest angular-momentum component of the ground state wavefunction in single-particle methods.  Additionally, this highest channel should be used as the local channel as it will generally be most similar to missing, yet-higher angular-momentum channels. 





\section{Acknowledgments} 
The authors would like to thank John Trail and Richard Needs for helpful discussions.  This research has been supported by the Cornell Center for Materials Research NSF-IGERT: A Graduate Traineeship in Materials for a Sustainable Future (DGE-0903653), by the National Science Foundation under Award Number CAREER DMR-1056587, and by the Energy Materials Center at Cornell (EMC2), funded by the U.S. Department of Energy, Office of Science, Office of Basic Energy Sciences under Award Number DE-SC0001086.  This research used computational resources of the Texas Advanced Computing Center under Contract Number TG-DMR050028N and of the Computation Center for Nanotechnology Innovation at Rensselaer Polytechnic Institute.

\bibliography{biblio}
\end{document}